\documentclass[12pt,preprint]{emulateapj}

\shorttitle{Origin of comet C/2016 R2 (PanSTARRS)'s peculiar composition}
\shortauthors{Mousis et al.}

\usepackage[T1]{fontenc}
\usepackage{graphicx}
\usepackage{etex}
\usepackage {amsmath}
\usepackage[squaren, Gray, cdot]{SIunits}
\usepackage{color}
\usepackage{lineno}

\begin{document}


\title{Cold traps of hypervolatiles in the protosolar nebula at the origin of comet C/2016 R2 (PanSTARRS)'s peculiar composition}


\author{Olivier Mousis\altaffilmark{1}, Artyom Aguichine\altaffilmark{1}, Alexis Bouquet\altaffilmark{2}, Jonathan I. Lunine\altaffilmark{3}, Gr\'egoire Danger\altaffilmark{2}, Kathleen E. Mandt\altaffilmark{4}, and Adrienn Luspay-Kuti\altaffilmark{4}}


\altaffiltext{1}{Aix Marseille Univ, CNRS, CNES, LAM, Marseille, France {\tt olivier.mousis@lam.fr}}
\altaffiltext{2}{Aix Marseille Univ, CNRS, PIIM, F-13013 Marseille, France}
\altaffiltext{3}{Department of Astronomy, Cornell University, Ithaca, NY 14853, USA}
\altaffiltext{4}{Applied Physics Laboratory, Johns Hopkins University, 11100 Johns Hopkins Rd., Laurel, MD 20723, USA}



\begin{abstract}
Recent observations of the long period comet C/2016 R2 (PanSTARRS) indicate an unusually high N$_2$/CO abundance ratio, typically larger than $\sim$0.05, and at least 2--3 times higher than the one measured in 67P/Churyumov-Gerasimenko. Another striking compositional feature of this comet is its heavy depletion in H$_2$O (H$_2$O/CO $\sim$0.32\%), compared to other comets. Here, we investigate the formation circumstances of a generic comet whose composition reproduces these two key features. We first envisage the possibility that this comet agglomerated from clathrates, but we find that such a scenario does not explain the observed low water abundance. We then alternatively investigate the possibility that the building blocks of the comet C/2016 R2 (PanSTARRS) agglomerated from grains and pebbles made of pure condensates via the use of a disk model describing the radial transport of volatiles. We show that N$_2$/CO ratios reproducing the value estimated in this comet can be found in grains condensed in the vicinity of the CO and N$_2$ icelines. Moreover, high CO/H$_2$O ratios (>~100~times the initial gas phase value) can be found in grains condensed in the vicinity of the CO iceline. If the building blocks of a comet assembled from such grains, they should present N$_2$/CO and CO/H$_2$O ratios consistent with the measurements made in comet C/2016 R2 (PanSTARRS)’s coma. Our scenario indicates that comet C/2016 R2 (PanSTARRS) formed in a colder environment than the other comets that share more usual compositions. Our model also explains the unusual composition of the interstellar comet 2l/Borisov.
\end{abstract}

\keywords{astrobiology -- comets: general -- comets: individual (C/2016 R2 (PanSTARRS)) -- methods: numerical -- solid state: volatile}

\section{Introduction}

Comets are supposed to be water-rich bodies that are relics of the formation of the solar system. Because they have undergone little alteration during the 4.6 billion years of the solar system evolution, investigating their composition provides indications of the physico-chemical conditions that were at play during their formation in the protosolar nebula (PSN). An outstanding question remaining to be solved is the origin of the apparent N$_2$ deficiency observed in comets while both Pluto and Triton, also formed in the outer solar system, harbor N$_2$--rich surfaces and atmospheres \citep{Le11,Cr15,Ma17}. Decades of remote-sensing observations of comets suggest they are depleted in N$_2$ \citep{Co00}. Only upper limits of $\sim$10$^{-4}$ for N$_2^+$/CO$^+$ were derived in the coma of comets 122P/1995 S1 (deVico) and C/1995 O1 (Hale-Bopp) \citep{Co00} while very few detections of N$_2^+$ emission lines have been reported in other comets from ground-based facilities. This apparent N$_2$ depletion was interpreted as the result of the selective trapping of CO at the expense of N$_2$ in the building blocks of comets presumably agglomerated from clathrates \citep{Ir03}, or as the result of their partial devolatilization due to radiogenic heating \citep{Mo12}. To a lesser extent, these important depletions are in agreement with the measurements of the N$_2$/CO ratio acquired in comet 67P/Churyumov-Gerasimenko (hereafter 67P/C-G) by the ROSINA mass spectrometer aboard the Rosetta spacecraft very early in the mission in October 2014, which suggest a value of (5.70 $\pm$ 0.66) $\times$ 10$^{-3}$ that is depleted by a factor of $\sim$25 as compared to the one derived from protosolar N and C abundances \citep{Ru15}. However, this N$_2$/CO ratio was measured at a heliocentric distance beyond 3 au, far beyond perihelion. \cite{Ru20} later derived a N$_2$/CO ratio of $\sim$2.87 $\times$ 10$^{-2}$ in 67P/C-G, a value obtained in May 2015 a few months before perihelion passage, and roughly 5 times larger than the previous one.

\begin{figure}
\resizebox{\hsize}{!}{\includegraphics[angle=0]{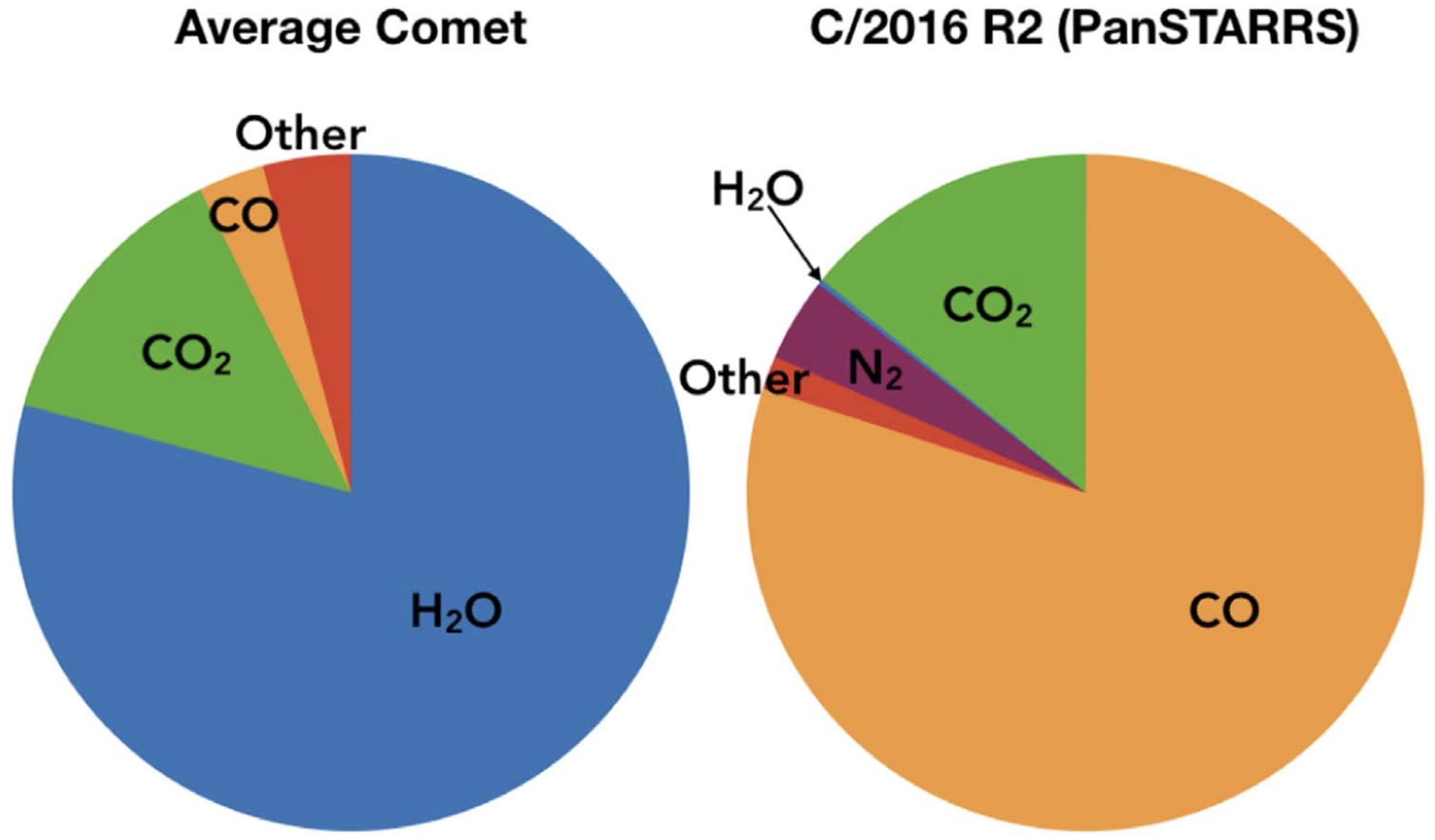}}
\caption{Comparison of the volatile composition of R2 to the average comet as derived by \citet{Mc19} [Reprinted from The Astronomical Journal, 158:128 (24pp), McKay, A., et al., The Peculiar Volatile Composition of CO-dominated Comet C/2016 R2 (PanSTARRS), with permission from IOP].
}
\label{fig1}
\end{figure}

Contrasting with most of these previous measurements, recent optical spectra of the long period comet C/2016 R2 (PanSTARRS) (hereafter R2) performed at a heliocentric distance of $\sim$3 AU, showed that its spectrum was largely dominated by the emission band of CO$^+$, but also surprisingly by the presence of the emission band of N$_2^+$ \citep{Co18a}. Subsequent observations confirmed an unusually high N$_2$/CO ratio in R2, a value estimated to range between 0.06 $\pm$ 0.01 \citep{Op19} and 0.08 \citep{Bi18}, and at least 2--3 times higher than the one measured closer to perihelion in 67P/C-G. Another striking compositional feature of this comet is its heavy depletion in H$_2$O (H$_2$O/CO $\sim$0.32\%; \cite{Mc19}). These trends are at odds with our understanding of the thermodynamic evolution of cometary nuclei which should be, in principle, depleted in the most volatile species instead of being enhanced, as R2 shows, after multiple solar passages. Since sublimation of supervolatile ices can occur possibly beyond 40 AU for CO \citep{Wo17}, one should expect CO outgassing, and then cometary activity over a significant fraction of R2's orbit. Figure \ref{fig1} represents a comparison of the volatile composition of R2 to the average comet made by \cite{Mc19}. The figure shows that all oxygen appears to be locked in CO and CO$_2$, and not H$_2$O, as is the case for typical comets.

In this paper, we aim at investigating the formation circumstances of a generic comet whose composition reproduces the two key features of R2's coma, namely its high N$_2$/CO ratio and the observed significant water depletion, assuming they reflect its primordial composition \citep{Mc19}. We first investigate the possibility that this comet could have agglomerated from clathrates as it has been suggested in the case of 67P/C-G \citep{Lu16,Mo16,Mo18}, but we find that such a scenario cannot explain the observed low water abundance. We then alternatively investigate the possibility that the building blocks of this comet agglomerated from grains and pebbles made of pure condensates crystallized in the vicinity of the CO iceline in the PSN. This scenario matches both observed features and suggests that this comet formed in a colder environment than the other comets sharing usual compositions.

\section{Agglomeration from clathrates?}
\label{clat}

For decades, the agglomeration of comets from clathrates has been regularly invoked to explain their compositional and thermodynamic properties \citep{Lu85,Kl86,Ir03,Ma10,Ma11,Ma12}. Recent measurements of 67P/C-G's composition by the Rosina mass spectrometer aboard the {\it Rosetta} mission have also been interpreted in favor of the presence of clathrates in its interior \citep{Lu16,Mo16,Mo18}. This possibility is now investigated in the case of R2, which presents unusually high CO and N$_2$ abundances in its coma, compared to typical comets. To do so, we aim at comparing the plausible composition of multiple guest (MG) clathrates formed in the PSN to that of R2, assuming this latter would have agglomerated from these icy structures. These MG clathrates are assumed to be formed from a gaseous mixture composed of N$_2$ and CO, assuming these species are the two main C-- and N--bearing volatiles in the gas phase of the disk, a hypothesis consistent with the thermochemical models \citep{Le80,Pr89,Mo02}. Because CO has a much higher propensity for clathration than N$_2$, the calculations utilized in this work are those performed in the framework of structure I clathrates, namely the structure predicted for a CO--dominated clathrate \citep{Mo05}. Given the fact that half of protosolar carbon is expected to be used in organic compounds in the PSN \citep{Po94}, only the remaining half is assumed to be in the form of CO. Assuming that 90\% of protosolar nitrogen is in N$_2$ form \citep{Le80}, we derive a N$_2$/CO ratio of 2.66~$\times$~10$^{-1}$ in the initial gas phase of the PSN, based on the protosolar abundances given by \cite{Lo09}. In our model, clathrates form in the PSN as long as crystalline water is available. Here we assume that once all the water budget has been used for clathration, the remaining volatiles are not incorporated into solids. 

Our model assumes the formation of a multiple guest (MG) clathrate with an equilibrium pressure expressed as \citep{Lu85}:

\begin{equation}
P_{eq,MG} = \left [\sum_i \frac{y_i}{P_{eq,i}} \right ]^{-1},
\end{equation}

\noindent where $y_i$ is the mole fraction of the component $i$ in the fluid phase. The equilibrium pressure curves of each species are determined by fitting the available theoretical and laboratory data \citep{Lu85} with equations of the form $ \log P_{eq,i} = A/T + B$, where $P_{eq,i}$ and $T$ are the partial equilibrium pressure and temperature of the considered species $i$, respectively. The relative abundances of guest species incorporated in MG clathrate formed at a given temperature and pressure from the PSN gas phase are calculated following the method described in \cite{Lu85} and \cite{Mo10}. This formalism has been used to interpret 67P/C-G's ice structure and composition from Rosetta/ROSINA observations of its coma \citep{Mo16,Mo18}.

\begin{figure}
\resizebox{\hsize}{!}{\includegraphics[angle=0]{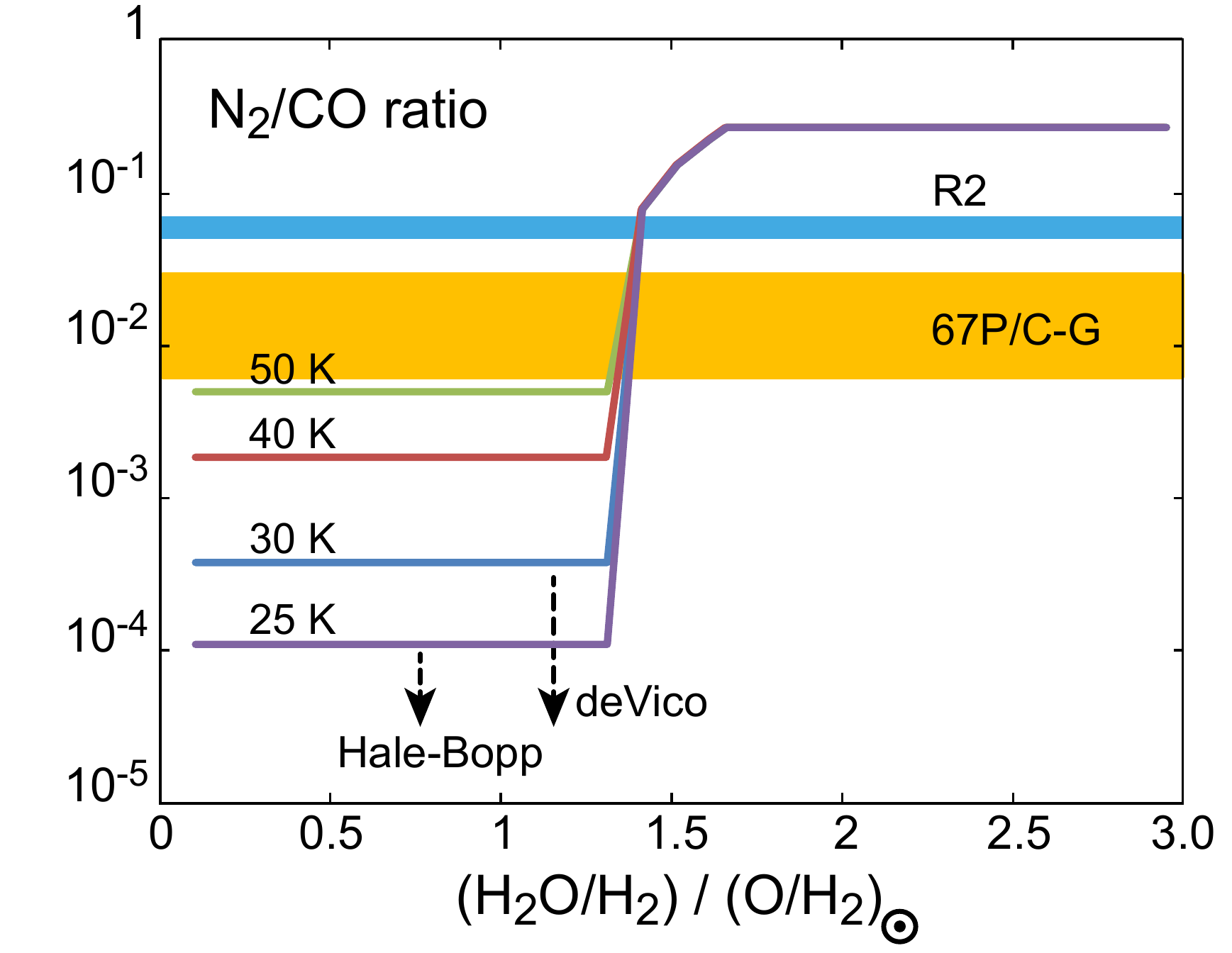}}
\caption{N$_2$/CO ratio in cometary grains calculated as a function of the water abundance (normalized to the oxygen protosolar abundance \citep{Lo09}) in the PSN for different formation temperatures. The blue and orange horizontal bars represent the N$_2$/CO ratios measured in comet R2 \citep{Op19} and 67P/C-G \citep{Ru15,Ru20}, respectively. The two vertical dashed lines with arrows down correspond to upper limits of the N$_2$/CO ratio measured in 122P/1995 S1 (deVico) and C/1995 O1 (Hale-Bopp) \citep{Co00}.}
\label{fig2}
\end{figure}

Figure \ref{fig2} shows the influence of the abundance of crystalline water ice in the outer PSN on the N$_2$/CO ratio in cometary grains at clathration temperatures of 25, 30, 40, and 50 K. These temperatures are all above the formation temperature of N$_2$ and CO pure condensates. For water abundances below 1.3~$\times$~(O/H$_2$)$_\odot$, the N$_2$/CO ratios in icy grains correspond to those calculated in clathrates at specified formation temperatures. For example, the N$_2$/CO ratios measured in Hale-Bopp, deVico and 67P/C-G are closely matched at 25, 30, and 50 K, respectively. At higher water abundances, CO is fully entrapped and significant amounts of N$_2$ become progressively trapped in clathrates, implying an increase of the global N$_2$/CO in icy grains. A water abundance of $\sim$1.4 $\times$ (O/H$_2$)$_\odot$ in the PSN allows one to retrieve a N$_2$/CO ratio in clathrates corresponding to the value measured in R2. Water abundances greater than $\sim$1.7 $\times$ (O/H$_2$)$_\odot$ allows the full trapping of the available N$_2$ in clathrates crystallized in the PSN, leading to a N$_2$/CO ratio of $\sim$2.66~$\times$~10$^{-1}$ in cometary grains, corresponding to the value calculated from protosolar N and C (with half C in CO and 90\% N in N$_2$). These calculations illustrate the fact that, in principle, a slight increase of the water abundance in the PSN can lead to the full trapping of CO and enough N$_2$ to reproduce the N$_2$/CO ratio in R2. 

However, the amount of water required by clathrates formed in the PSN to match the N$_2$/CO ratio measured in R2's coma is inconsistent with its estimated low abundance. To match the N$_2$/CO measured in R2, our clathrate model requires a H$_2$O/CO ratio of 6.1. In contrast, the H$_2$O/CO mixing ratio measured at 2.8 AU in R2 is $\sim$0.32\%, a value several orders of magnitude lower than those measured in 67P/C-G ($\sim$200--1900\%) at a similar distance range \citep{Ga17} or in many other comets \citep{Mc19}. As a result, this mechanism fails at challenging to explain the H$_2$O depletion in R2-like comets via this mechanism.

\section{Agglomeration from particles formed in the vicinity of CO and N$_2$ ice lines}

Disk models including radial transport of volatiles can display significant enrichments of both volatile and refractory matter at its condensation location, implying potentially drastic changes of the local metallicity in the disk \citep{St88,Cy99,Al14,Mo19,Mo20,Ag20}. Here we investigate this effect by considering the possibility that R2 formed in the vicinity of the condensation locations of CO and N$_2$ crystalline ices in the PSN. 

\subsection{Model}

The volatile transport and distribution model used in our work is the one described in \cite{Ag20} and \cite{Mo20}, to which the reader is referred for details. Three energy sources are considered in our model, namely viscous heating, irradiation from the current Sun, and ambient constant irradiation giving a background temperature of 10 K (see Eq. 6 of \cite{Ag20}). Here, the accretion rate onto the star declines over time, and the disk develops a transition radius between inward and outward gas flows. The location of this transition radius is moving outward with time. In a few words, our time-dependent PSN model is governed by the following differential equation \citep{Ly74}:

\begin{eqnarray}
\frac{\partial \Sigma_{\mathrm{g}}}{\partial t} = \frac{3}{r} \frac{\partial}{\partial r} \left[ r^{1/2} \frac{\partial}{\partial r} \left( r^{1/2} \Sigma_{\mathrm{g}} \nu \right)\right],
\label{eqofmotion}
\end{eqnarray}

\noindent This equation describes the time evolution of a viscous accretion disk of surface density $\Sigma_{\mathrm{g}}$ of viscosity $\nu$, assuming hydrostatic equilibrium in the $z$ direction. The viscosity is calculated in the framework of the $\alpha$-formalism \citep{Sh73} using the following method. For each distance $r$ to the Sun, the disk's properties are calculated by solving the equation of energy balance between viscous heating and radiative transfer at the midplane level. This gives us $\nu$, as well as the pressure and temperature profiles of the disk as a function of $r$. The evolution of the disk starts with an initial profile given by $\Sigma_{\mathrm{g}} \nu \propto \exp \left(-r^{2-p}\right)$, with $p=\frac{3}{2}$ for an early disk \citep{Ly74}. In our computations, the initial disk's mass is fixed to 0.1~M$_\odot$. The computational box is set equal to 500 AU, allowing 99\% of the disk mass to be encapsulated within $\sim$100 AU. Mass can only be lost by accretion onto the Sun or via outward diffusion, the boundary between these two regimes being defined by location of  $R_c$, the centrifugal radius. $R_c$ is initially located at $\sim$6 AU and expands up to $\sim$20 and 45 AU after 0.1 and 1 Myr of evolution, respectively. The disk's initial mass accretion rate onto the Sun is set to $10^{-7.6}$~M$_\odot$~yr$^{-1}$ \citep{Ha98}. 

The size of dust particles used in our model is determined by a two-populations algorithm derived from \cite{Bi12}. This algorithm computes the representative size of particles through the estimate of the limiting Stokes number in various dynamical regimes. In our model, dust is initially present in the form of particles of sizes $a_0 = 10^{-7}$ m, and grow through mutual collisions. This growth is limited by the maximum sizes imposed by fragmentation or by the drift velocity of the grains (see \cite{Ag20} for details). The dust surface density is the sum over all surface densities of available solids at given time and location, assuming a protosolar ice-to-rock ratio ($\sim$1.58) \citep{Lo09}, and a bulk density of 1000 kg m$^{-3}$.

We follow the approaches of \cite{De17} and \cite{Dr17} for the dynamics of trace species in terms of motion and thermodynamics, respectively. We assume that the disk is uniformly filled with H$_2$O, CO, and N$_2$. The abundances of CO, and N$_2$ are those derived in Sec. \ref{clat} and the abundance of H$_2$O is that of the leftover oxygen. No chemistry is assumed to happen between the trace species. In our simulations, grains are considered as homogeneous mixtures of solid species in proportions determined from the aforementioned assumptions (relative abundance ratios, condensation, sublimation). Sublimation of grains occurs during their inward drift when partial pressures of trace species become lower than the corresponding vapor pressures. Once released, vapors diffuse both inward and outward. Because of the outward diffusion, vapors can recondense back in solid form following the rates defined by \cite{Dr17}, and condensation occurs either until thermodynamic equilibrium is reached or until no more gas is available to condense. The position of the iceline of a given species is defined as the location where its gaseous abundance equals that of its coexisting solid \citep{Lo03,Ob19,Ag20}. This approach results in icelines closer to the Sun than those computed via a simple comparison between the partial and saturation pressures.

The motion of dust and vapor is computed by integrating the 1D radial advection-diffusion equation derived from \cite{Bi12} and \cite{De17}, and detailed in \cite{Ag20}. The vapor pressures of trace species are taken from \cite{Fr09}.

\subsection{Results}

\begin{figure*}
\resizebox{\hsize}{!}{\includegraphics[angle=0]{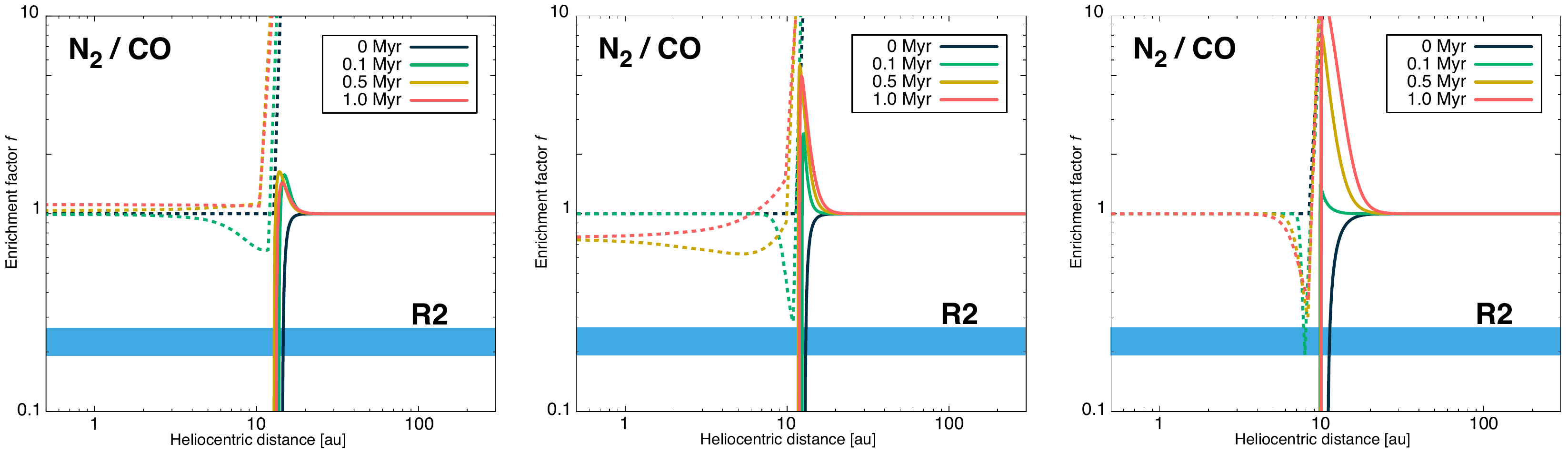}}
\caption{From left to right: radial profiles of the N$_2$/CO ratio relative to its initial abundance (defined by the enrichment factor $f$) calculated as a function of time in the PSN for viscosity parameters $\alpha$~=~ 5 $\times$ 10$^{-3}$, 10$^{-3}$, and 10$^{-4}$. Dashed and solid lines correspond to vapor and solid phases, respectively. The blue bar corresponds to the N$_2$/CO ratio measured in comet R2.}
\label{fig3}
\end{figure*}

Figure \ref{fig3} represents the time evolution of the radial profiles of the N$_2$/CO ratio relative to its assumed initial ratio ($\sim$2.66~$\times$~10$^{-1}$), defined by the enrichment factor $f$, in both solid and gas phases of the PSN, and for three values of the viscosity parameter $\alpha$, namely 5 $\times$ 10$^{-3}$, 10$^{-3}$, and 10$^{-4}$. The adopted $\alpha$ values are well within the range of those typically used in models of protoplanetary disks \citep{He01,Ne13,Si15}. The figure shows that $f$ is flat almost everywhere in the PSN, except in the vicinity of the CO and N$_2$ icelines, which are located at very close distances from each other, i.e. less than a few tenths of AU, in the 10--15 AU region of the PSN, with the CO iceline situated a bit closer in in the disk.

Depending on the choice of the $\alpha$ parameter, the N$_2$/CO ratio increases to more than $\sim$10 in the gas phase in the area of the two icelines. This peak corresponds to the supply of N$_2$ vapor when N$_2$-rich dust drifts inward the N$_2$ iceline, which is in excess compared to the CO vapor supplied via backward diffusion beyond the CO iceline. When significant, this peak is preceded by a decrease of the N$_2$/CO, which corresponds to the the supply of CO vapor when CO-rich dust drifts inward the CO iceline. The N$_2$/CO ratio in solid phase also experiences important depletions, which correspond to the location where N$_2$ essentially forms vapor while CO remains in solid phase. The N$_2$/CO ratio in solid phase also reaches peaks up to $f$ $\sim$10, depending on the $\alpha$ value, at the location of the N$_2$ iceline, corresponding to the formation of solid N$_2$ from N$_2$ vapor diffusing backward its corresponding iceline. The figure also shows it is possible to form dust in a narrow region of the PSN, i.e. within 10--15 AU, with N$_2$/CO ratios matching the value estimated in comet R2. A comet agglomerated from these grains would present a N$_2$/CO ratio consistent with the R2 value, both in its interior and its coma, because the two molecules present similar volatilities, as testified by the proximity of their icelines in the PSN.

\begin{figure*}
\resizebox{\hsize}{!}{\includegraphics[angle=0]{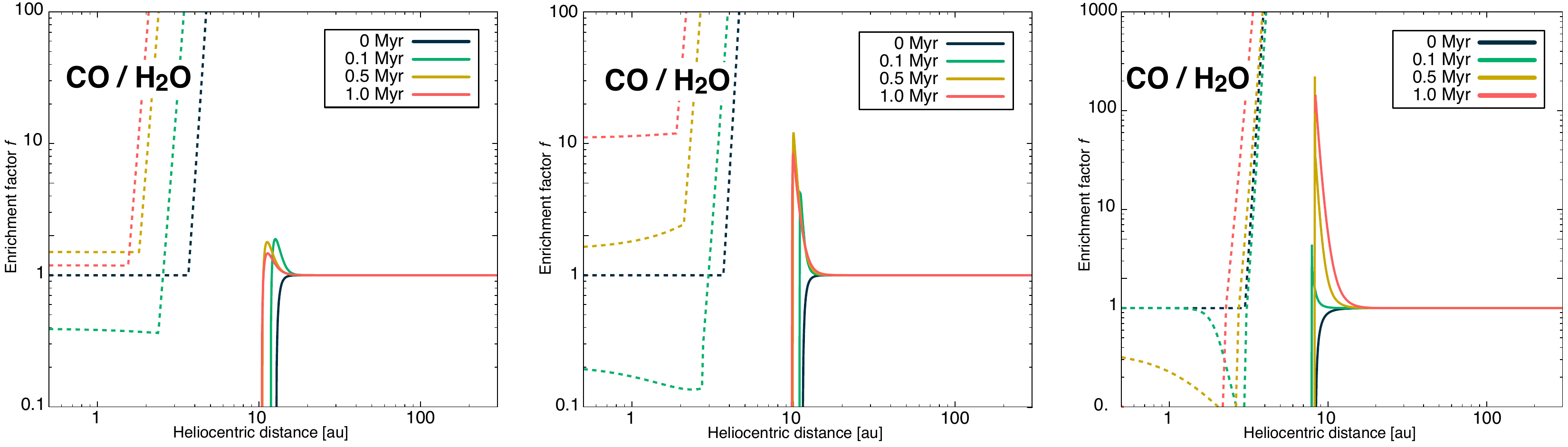}}
\caption{From left to right: radial profiles of the CO/H$_2$O ratio relative to its initial abundance (defined by the enrichment factor $f$) calculated as a function of time in the PSN for viscosity parameters $\alpha$~=~5 $\times$ 10$^{-3}$, 10$^{-3}$, and 10$^{-4}$. Dashed and solid lines correspond to vapor and solid phases, respectively.}
\label{fig4}
\end{figure*}

Figure \ref{fig4} represents the time evolution of the radial profiles of the CO/H$_2$O ratio relative to its assumed initial ratio ($\sim$2.96~$\times$~10$^{-1}$), defined by the enrichment factor $f$, in both solid and gas phases of the PSN, and for three values of the viscosity parameter $\alpha$, namely 5~$\times$~10$^{-3}$, 10$^{-3}$, and 10$^{-4}$. The CO/H$_2$O ratio in the gas phase increases steeply in the vicinity of the H$_2$O iceline, which is located at much closer distances from the Sun than the CO and N$_2$ icelines. This steep increase is due to the freezing of the H$_2$O vapor when it diffuses outward through its iceline. The CO/H$_2$O ratio is deeply depleted in the solid phase at distances inward of the CO iceline because this species remains in gaseous form. On the other hand, when approaching the CO iceline, the CO/H$_2$O ratio presents values exceeding its initial value. A peak forms at this location, and its magnitude depends on the adopted viscosity parameter $\alpha$ of the disk. The CO/H$_2$O enrichment factor $f$ can typically reach a value of $\sim$10 with $\alpha$~=~10$^{-3}$, while it peaks at $\sim$200 with $\alpha$~=~10$^{-4}$. If the building blocks of a comet were assembled from grains formed at this location of the PSN, they should present a CO--rich and H$_2$O--poor composition. Assuming $f$ = 100, the CO/H$_2$O ratio would be $\sim$30, a value exceeding by far those known in typical comets \citep{Bo17}. We note that this value is still about 10 times lower than the one inferred in R2's coma but this latter can be seen as an overestimate of the bulk CO abundance in the comet. Indeed, CO is much more volatile than H$_2$O and its vapor comes from deeper layers than H$_2$O vapor in the nucleus, potentially inducing an enriched CO/H$_2$O ratio in the coma. This effect has been investigated in depth by \cite{Ma14} who found that the abundance of volatile molecules (relative to H$_2$O) released from the interiors of nuclei can vary by several orders of magnitude along the comet's orbital evolution.

Interestingly, Figures \ref{fig3} and \ref{fig4} show that the locations of the CO and N$_2$ icelines do not significantly evolve with time and are only weakly affected by the value adopted for the viscosity parameter. The main reason for this is that viscous heating occurs mainly in the dense regions of the PSN. With our parametrization, the gas density quickly decreases after 10 AU, implying that the temperature profile beyond this distance is mostly ruled by solar irradiation, here assumed to be constant.

\section{Conclusions}

In this work, we first show that a comet agglomerated from clathrates crystallized in the PSN could, in principle, display N$_2$/CO ratios corresponding to the value measured in R2's coma. This finding contrasts with the suggestion of \cite{Wi18} to rule out the presence of clathrates in R2 only because its N$_2$/CO ratio is higher than those measured in other comets. This statement is correct in the case of a limited budget of available crystalline water in the PSN. However, if water is abundant enough, as shown in this paper, both molecules are efficient clathrate formers. On the other hand, the amount of water required by clathrates formed in the PSN to match the N$_2$/CO ratio measured in R2's coma, namely H$_2$O/CO $\sim$6, is inconsistent with its estimated extremely low value ($\sim$0.32\%). 

We have then used a disk model describing the radial transport of volatiles in the PSN, evolving both in gaseous and solid (pure condensates) phases. We show that N$_2$/CO ratios reproducing the value estimated in comet R2 can be found in dust formed in the vicinity of the CO and N$_2$ icelines, i.e. within the 10--15 AU region of the PSN, depending on the adopted viscosity parameter $\alpha$ parameter of the disk. Meanwhile, very high CO/H$_2$O ratios, i.e. up to more than 200 times the value derived from a protosolar gaseous mixture, can be found in dust formed in the vicinity of the CO iceline ($\sim$10 AU), the extent of which also depending on the adopted viscosity parameter. If the building blocks of a comet assembled from grains formed close to the CO iceline in the PSN, they should present N$_2$/CO and CO/H$_2$O ratios consistent with the measurements made in comet R2's coma. 

\begin{figure}
\resizebox{\hsize}{!}{\includegraphics[angle=0]{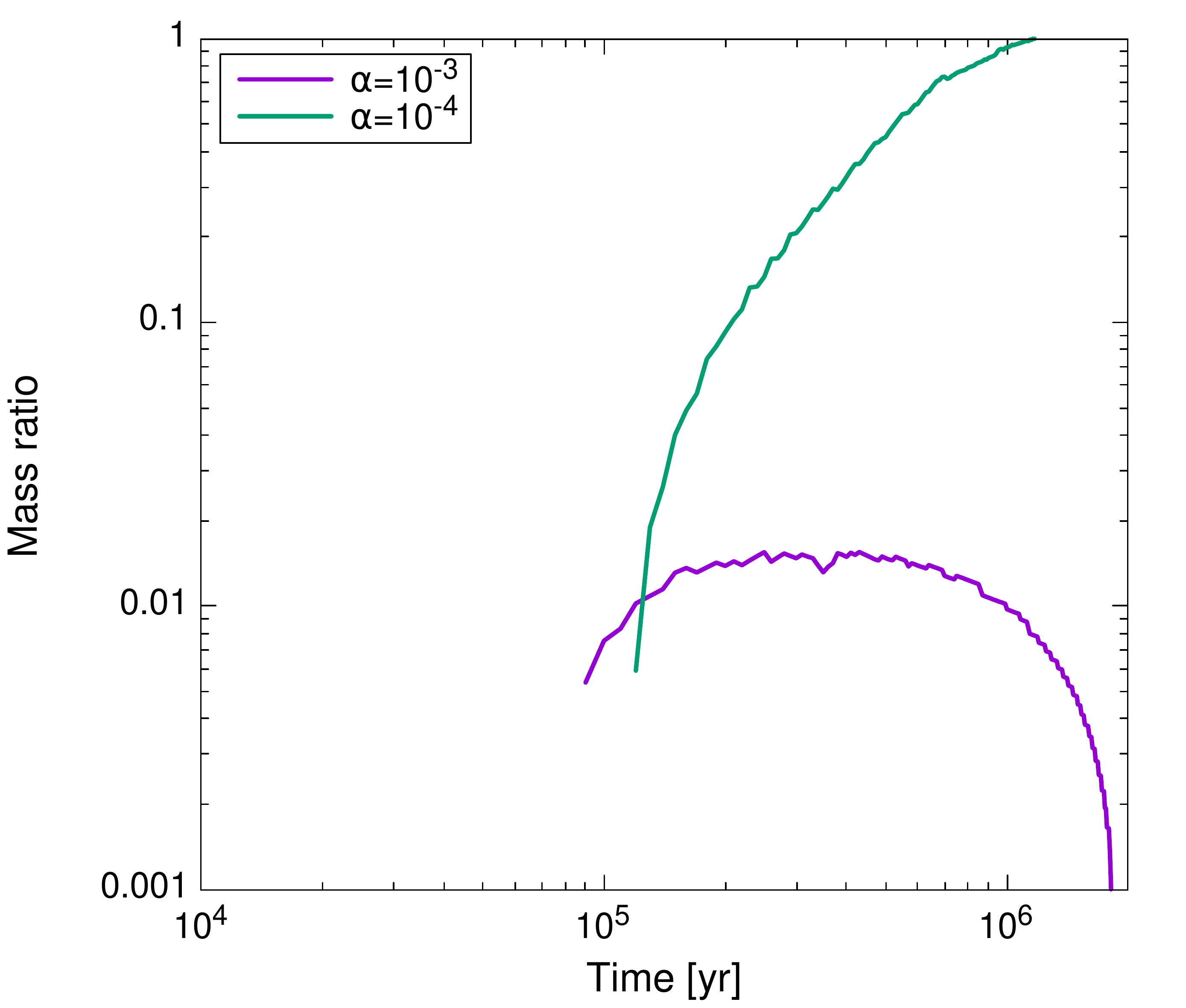}}
\caption{Time evolution of the mass ratio between reservoir 1 (R2-like comets) and reservoir 2 (average comets) for viscosity parameters $\alpha$ = 10$^{-3}$ and 10$^{-4}$.}
\label{fig5}
\end{figure}

The likelihood of finding R2-like comets versus H$_2$O-rich comets in the PSN can be, to first order, assessed via comparing the masses of their respective reservoirs. To do so, we calculated the mass of H$_2$O and CO ices contained in a PSN annulus in which CO was found to be more abundant than H$_2$O (R2-like comets -- reservoir 1) , and in another annulus where the CO/H$_2$O ratio was set to 2.96 $\times$~10$^{-1}$, i.e. the value derived from protosolar O and C abundances (average comets -- reservoir 2). The time evolution of the reservoir 1/reservoir 2 mass ratio is represented in Fig. \ref{fig5} for $\alpha$ = 10$^{-3}$ and 10$^{-4}$. Higher $\alpha$ values do not allow for the formation of reservoir 1 in the PSN. This is also shown by Fig. \ref{fig4} in the case $\alpha$~=~5~$\times$~10$^{-3}$, where the CO/H$_2$O ratio in the solid phase never exceeds twice the initial ratio. Figure 5 illustrates the fact that the mass ratio between the two reservoirs strongly depends on the adopted $\alpha$ parameter and the epoch of the PSN evolution. In the case $\alpha$ = 10$^{-3}$, the mass of R2-like comets is about 1\% that of the average comets, if the grains from which they assembled condensed between 0.1 and 1 Myr in the PSN. In the case $\alpha$ = 10$^{-4}$, the mass of R2-like comets varies between less than $\sim$1\% and 100$\%$ that of the average comets, depending on the particular grain formation epoch in the 0.1--1 Myr timeframe. We conclude that the likelihood of finding R2-like comets versus H$_2$O-rich comets in the PSN strongly depends on the adopted disk parameters, and thus remains difficult to predict a priori.

The heliocentric distances indicated for the different icelines must be taken with caution and are model-dependent. However, the agglomeration of R2 from dust condensed in the region of the CO iceline indicates that this comet formed at a greater heliocentric distance than a H$_2$O-rich comet formed from clathrates because the formation temperatures of N$_2$-- and CO--rich clathrates are always higher than those of the pure condensates at PSN conditions. Our model also applies to other protoplanetary disks and can explain the unusual composition of the interstellar comet 2l/Borisov, which seems to contain substantially more CO than H$_2$O gas in its coma \citep{Bo20}. 

Interestingly, physicochemical models of nitrogen chemistry in protostellar disks show that photodissociation of N$_2$ leads to production of HCN \citep{Hi17}. To overcome this issue, R2 may have formed in a location of the PSN where significant N$_2$ shielding led to the high N$_2$ and decreased HCN abundances \citep{Wi18}.



\acknowledgements
We thank the two Referees for their valuable comments on the content of our manuscript and their suggestions for improving the document. O.M. and A.B. acknowledge support from CNES, and J.L. from the JWST project. K.E.M. acknowledges support from the NASA RDAP grant 80NSSC19K1306, and A.L-K. acknowledges support from the NASA RDAP grant 80NSSC18K1620.



\end{document}